%% file: STARJpsiPolarization.tex
\documentclass[aps,prd,letterpaper,preprintnumbers,superscriptaddress,showpacs,amsmath,letterpaper,floatfix,hyperref,nofootinbib,twocolumn]{revtex4}
\usepackage{multirow}
\usepackage{color}
\usepackage{tabularx}
\usepackage{graphicx}
\usepackage{amsthm}
\usepackage{amsmath}
\usepackage{amssymb}
\usepackage{bm}
\usepackage{textcomp}
\usepackage{url}
\usepackage{tabularx}
\usepackage{verbatim}

\newcommand{\pt}{$p_{T}$ }
\newcommand{\pT}{$p_{T}$ }
\newcommand{\Jpsi}{$J/\psi$ }

\begin{document}

\title{\boldsymbol{$J/\psi$} polarization in \boldsymbol{$p+p$} collisions\\at \boldsymbol{$\sqrt{s}$} = 200 GeV in STAR}

\input{authors.tex}

\pagenumbering{arabic}
\pagestyle{myheadings}
\thanks{}
\pacs{25.75.Dw, 13.20.Gd, 14.40.Lb}
\markboth{{\small STAR Collaboration}}{{\small $J/\psi$ polarization in $p+p$ collisions\\at $\sqrt{s}$ = 200 GeV in STAR}} 

\begin{abstract}

We report on a polarization measurement of inclusive $J/\psi$ mesons in the di-electron decay channel at mid-rapidity at 2 $<p_{T}<$ 6 GeV/$c$ in $p+p$ collisions at $\sqrt{s}$ = 200 GeV. Data were taken with the STAR detector at RHIC. The $J/\psi$ polarization measurement should help to distinguish between different models of the $J/\psi$ production mechanism since they predict different $p_{T}$ dependences of the $J/\psi$ polarization. In this analysis, $J/\psi$ polarization is studied in the helicity frame. The polarization parameter $\lambda_{\theta}$ measured at RHIC becomes smaller towards high $p_{T}$, indicating more longitudinal $J/\psi$ polarization as $p_{T}$ increases. The result is compared with predictions of presently available models. 

\end{abstract}

\maketitle

\input{Introduction.tex}

\input{Data.tex}
\input{Result.tex}
\input{Summary.tex}
\input{Acknowledgements.tex}

\bibliographystyle{model1a-num-names}
\bibliography{/home/barbara/Work/Bibliography_all}

\end{document}

%% file: authors.tex
\affiliation{AGH University of Science and Technology, Cracow, Poland}
\affiliation{Argonne National Laboratory, Argonne, Illinois 60439, USA}
\affiliation{University of Birmingham, Birmingham, United Kingdom}
\affiliation{Brookhaven National Laboratory, Upton, New York 11973, USA}
\affiliation{University of California, Berkeley, California 94720, USA}
\affiliation{University of California, Davis, California 95616, USA}
\affiliation{University of California, Los Angeles, California 90095, USA}
\affiliation{Universidade Estadual de Campinas, Sao Paulo, Brazil}
\affiliation{Central China Normal University (HZNU), Wuhan 430079, China}
\affiliation{University of Illinois at Chicago, Chicago, Illinois 60607, USA}
\affiliation{Cracow University of Technology, Cracow, Poland}
\affiliation{Creighton University, Omaha, Nebraska 68178, USA}
\affiliation{Czech Technical University in Prague, FNSPE, Prague, 115 19, Czech Republic}
\affiliation{Nuclear Physics Institute AS CR, 250 68 \v{R}e\v{z}/Prague, Czech Republic}
\affiliation{Frankfurt Institute for Advanced Studies FIAS, Germany}
\affiliation{Institute of Physics, Bhubaneswar 751005, India}
\affiliation{Indian Institute of Technology, Mumbai, India}
\affiliation{Indiana University, Bloomington, Indiana 47408, USA}
\affiliation{Alikhanov Institute for Theoretical and Experimental Physics, Moscow, Russia}
\affiliation{University of Jammu, Jammu 180001, India}
\affiliation{Joint Institute for Nuclear Research, Dubna, 141 980, Russia}
\affiliation{Kent State University, Kent, Ohio 44242, USA}
\affiliation{University of Kentucky, Lexington, Kentucky, 40506-0055, USA}
\affiliation{Korea Institute of Science and Technology Information, Daejeon, Korea}
\affiliation{Institute of Modern Physics, Lanzhou, China}
\affiliation{Lawrence Berkeley National Laboratory, Berkeley, California 94720, USA}
\affiliation{Massachusetts Institute of Technology, Cambridge, MA 02139-4307, USA}
\affiliation{Max-Planck-Institut f\"ur Physik, Munich, Germany}
\affiliation{Michigan State University, East Lansing, Michigan 48824, USA}
\affiliation{Moscow Engineering Physics Institute, Moscow Russia}
\affiliation{National Institute of Science Education and Research, Bhubaneswar 751005, India}
\affiliation{Ohio State University, Columbus, Ohio 43210, USA}
\affiliation{Old Dominion University, Norfolk, VA, 23529, USA}
\affiliation{Institute of Nuclear Physics PAN, Cracow, Poland}
\affiliation{Panjab University, Chandigarh 160014, India}
\affiliation{Pennsylvania State University, University Park, Pennsylvania 16802, USA}
\affiliation{Institute of High Energy Physics, Protvino, Russia}
\affiliation{Purdue University, West Lafayette, Indiana 47907, USA}
\affiliation{Pusan National University, Pusan, Republic of Korea}
\affiliation{University of Rajasthan, Jaipur 302004, India}
\affiliation{Rice University, Houston, Texas 77251, USA}
\affiliation{Universidade de Sao Paulo, Sao Paulo, Brazil}
\affiliation{University of Science \& Technology of China, Hefei 230026, China}
\affiliation{Shandong University, Jinan, Shandong 250100, China}
\affiliation{Shanghai Institute of Applied Physics, Shanghai 201800, China}
\affiliation{SUBATECH, Nantes, France}
\affiliation{Temple University, Philadelphia, Pennsylvania, 19122, USA}
\affiliation{Texas A\&M University, College Station, Texas 77843, USA}
\affiliation{University of Texas, Austin, Texas 78712, USA}
\affiliation{University of Houston, Houston, TX, 77204, USA}
\affiliation{Tsinghua University, Beijing 100084, China}
\affiliation{United States Naval Academy, Annapolis, MD 21402, USA}
\affiliation{Valparaiso University, Valparaiso, Indiana 46383, USA}
\affiliation{Variable Energy Cyclotron Centre, Kolkata 700064, India}
\affiliation{Warsaw University of Technology, Warsaw, Poland}
\affiliation{University of Washington, Seattle, Washington 98195, USA}
\affiliation{Yale University, New Haven, Connecticut 06520, USA}
\affiliation{University of Zagreb, Zagreb, HR-10002, Croatia}

\author{L.~Adamczyk}\affiliation{AGH University of Science and Technology, Cracow, Poland}
\author{J.~K.~Adkins}\affiliation{University of Kentucky, Lexington, Kentucky, 40506-0055, USA}
\author{G.~Agakishiev}\affiliation{Joint Institute for Nuclear Research, Dubna, 141 980, Russia}
\author{M.~M.~Aggarwal}\affiliation{Panjab University, Chandigarh 160014, India}
\author{Z.~Ahammed}\affiliation{Variable Energy Cyclotron Centre, Kolkata 700064, India}
\author{I.~Alekseev}\affiliation{Alikhanov Institute for Theoretical and Experimental Physics, Moscow, Russia}
\author{J.~Alford}\affiliation{Kent State University, Kent, Ohio 44242, USA}
\author{C.~D.~Anson}\affiliation{Ohio State University, Columbus, Ohio 43210, USA}
\author{A.~Aparin}\affiliation{Joint Institute for Nuclear Research, Dubna, 141 980, Russia}
\author{D.~Arkhipkin}\affiliation{Brookhaven National Laboratory, Upton, New York 11973, USA}
% ...

\collaboration{STAR Collaboration}\noaffiliation

%% file: Introduction.tex
\section[Introduction]{Introduction}

The \Jpsi is a bound state of charm ($c$) and anti-charm ($\overline{c}$) quarks. Charmonia physical states have to be colorless, however they can be formed via a color-singlet or a color-octet intermediate $c\overline{c}$ state. The first model of charmonia production, the Color Singlet Model (CSM) \cite{Einhorn:1975ua, Ellis:1976fj, Carlson:1976cd, Chang:1980, Berger:1980ni, Baier:1981uk, Baier:1981zz, Baier:1983va}, assumed that $c\overline{c}$ pairs are created in the color-singlet state only. This early prediction failed to describe the measured charmonia cross-section which has led to the development of new models. For example, Non-Relativistic QCD (NRQCD) \cite{Braaten:1996pv} calculations were proposed in which a $c\overline{c}$ color-octet intermediate state, in addition to a color-singlet state, can bind to form a charmonium. 

Different models of \Jpsi production are able to describe the measured \Jpsi production cross section reasonably well \cite{Abelev:2009qaa, Adamczyk:2012ey, Adare:2011vq, PhysRevLett.79.572, Acosta:2004yw, Aad:2011sp, Khachatryan:2010yr, Aaij:2011jh} and therefore other observables are needed to discriminate between different $J/\psi$ production mechanisms. $J/\psi$ spin alignment, commonly known as polarization, can be used for this purpose, since various models predict different transverse momentum ($p_{T}$) dependence for the polarization. The predictions of different models deviate the most at high $p_{T}$. Therefore a high-$p_{T}$ \Jpsi polarization measurement is of particular interest since it can help to discriminate between the models.

NRQCD calculations with color-octet contributions \cite{Braaten:1999qk} are in good agreement with observed \Jpsi \pT spectra in different experiments at different energies, at the Relativistic Heavy Ion Collider (RHIC) \cite{Adamczyk:2012ey,Adare:2011vq}, the Tevatron \cite{PhysRevLett.79.572, Acosta:2004yw} and the Large Hadron Collider (LHC) \cite{Khachatryan:2010yr, Aaij:2011jh, Butenschoen:2012qh}. But the calculations fail to describe the \Jpsi polarization at high $p_{T}$ ($p_{T} >$ 5 GeV/$c$) measured by the CDF experiment at FermiLab at $\sqrt{s}$ = 1.96 TeV \cite{PhysRevLett.99.132001}. NRQCD calculations predict transverse polarization for $p_{T} $ $>$ 5 GeV/$c$ and the growth of the polarization parameter $\lambda_{\theta}$ with increasing $p_{T}$ \cite{Butenschoen:2012px}. However, the CDF polarization measurement becomes slightly longitudinal with increasing $p_{T}$, for 5 $< p_{T} <$ 30 GeV/$c$ \cite{PhysRevLett.99.132001}. 
Also, the CMS \Jpsi polarization measurement in $p+p$ collisions at $\sqrt{s}$ = 7 TeV for high transverse momenta \cite{Chatrchyan:2013cla} is in disagreement with existing next-to-leading-order (NLO) NRQCD calculations \cite{Butenschoen:2012px,Gong:2012ug}. 
In addition, the \Jpsi polarization measurements at the same energy and for lower $p_{T}$ were performed by ALICE (inclusive \Jpsi production) \cite{Abelev:2011md} and LHCb (prompt \Jpsi production) \cite{Aaij:2013nlm} experiments at forward rapidity. The ALICE experiment observed zero polarization while LHCb $\lambda_{\theta}$ results indicate small longitudinal polarization (with other coefficients consistent with zero). Data from both experiments favor NLO NRQCD over NLO CSM \cite{Aaij:2013nlm, Butenschoen:2012px}.
At RHIC energies, at intermediate $p_{T}$ (1.5 $< p_{T}  \lesssim$ 5 GeV/$c$) and for mid-rapidity, the tuned leading-order (LO) NRQCD model \cite{PhysRevD.81.014020} predicts slightly longitudinal \Jpsi polarization and describes the PHENIX result \cite{PhysRevD.82.012001} well.

In the case of the Color Singlet Model, the Next-to-Leading Order calculations (NLO$^+$ CSM) \cite{Lansberg:2010vq_paper} for the \pT spectrum are in near agreement with the RHIC data at low and mid $p_{T}$ and these CSM calculations predict longitudinal \Jpsi polarization at intermediate $p_{T}$ (1.5 $< p_{T}  <$ 6 GeV/$c$) at mid-rapidity which is in agreement with the PHENIX result \cite{Lansberg:2010vq_paper}. At the Tevatron and LHC energies, the upper bound of NNLO* prediction \cite{Lansberg:2011hi} is very close to the experimental cross section data, similar to RHIC \cite{Lansberg:2010vq_paper}. Also, the upper edge of this prediction for the polarization is in good agreement with the CDF data \cite{Lansberg:2011hi}. However, NLO CSM calculations \cite{Aaij:2013nlm, Butenschoen:2012px} do not describe \Jpsi polarization results from ALICE and LHCb well.

For the lower $p_{T}$ range at RHIC energies, the LO NRQCD calculations \cite{PhysRevD.81.014020} and NLO$^+$ CSM \cite{Lansberg:2010vq_paper}  have similar predictions regarding the \Jpsi polarization, which is longitudinal, and describe the experimental results\cite{PhysRevD.82.012001} well. However, these models predict different \pT dependence: in the case of the NRQCD prediction, the trend is towards the transverse polarization with increasing $p_{T}$, while the NLO$^+$ CSM shows almost no \pT dependence. Thus, it is especially important to measure a $p_{T}$ dependence of the \Jpsi polarization and go to high $p_{T}$.


In this paper, we report a \Jpsi polarization measurement in $p+p$ collisions at $\sqrt{s}$ = 200 GeV at rapidity (y) $|y| <$ 1, in the \pT range 2 $<p_{T}<$ 6 GeV/$c$ from the STAR experiment at RHIC. The analysis is done using data with a high-$p_{T}$ electron (so-called High Tower) trigger. The \Jpsi is reconstructed via its di-electron decay channel. The angular distribution parameter (polarization parameter) $\lambda_{\theta}$ for electron decay of the \Jpsi is extracted in the helicity frame \cite{PhysRevD.18.2447} as a function of $J/\psi$ $p_{T}$, in three \pT bins. The obtained result is compared with predictions of NLO$^+$ CSM \cite{Lansberg:2010vq_paper} and LO NRQCD calculations (COM) \cite{PhysRevD.81.014020}. 

\subsection{Angular distribution of decay products}

$J/\psi$ polarization is analyzed via the angular distribution of the decay electrons in the helicity frame \cite{PhysRevD.18.2447}. In this analysis, we are interested in the polar angle $\theta$. It is the angle between the positron momentum vector in the $J/\psi$ rest frame and the \Jpsi momentum vector in the laboratory frame.
The full angular distribution, which is derived from the density matrix elements of the production amplitude using parity conservation rules, is described by:
\begin{equation}
\frac{d^{2}N}{d(\cos\theta)d\phi} \propto 1+\lambda_\theta \cos^2\theta + \\ \lambda_\phi \sin^2\theta \cos2\phi + \lambda_{\theta\phi}\sin2\theta \cos\phi \,,
\label{eq:angularDistribution}
\end{equation}
where $\theta$ and $\phi$ are polar and azimuthal angles, respectively; $\lambda_\theta$ and $\lambda_\phi$ are the angular decay coefficients.
The angular distribution integrated over the azimuthal angle is parametrized as
\begin{equation}
\frac{dN}{d(\cos\theta)} \propto 1+\lambda_{\theta} \cos^2\theta \,,
\label{eq:angularDistribution_theta}
\end{equation}
where $\lambda_{\theta}$ is called the polarization parameter. This parameter contains both the longitudinal and transverse components of the  $J/\psi$ cross section; $\lambda_{\theta} = 1$ indicates full transverse polarization, and $\lambda_{\theta} = -1$ corresponds to full longitudinal polarization.

The measurement presented in this Letter is limited to the $\theta$ angle analysis due to statistical limitations. Extraction of the $\lambda_{\theta}$ parameter in the helicity frame allows one to compare the result with the available model predictions and draw model dependent conclusions. A measurement of the $\theta$ angle with a better precision, as well as the $\phi$ angle, will be possible with a newer STAR data at $\sqrt{s}$ = 500 GeV. Then, the frame invariant parameter, also in different reference frames, can be calculated providing model independent information about the $J/\psi$ polarization \cite{Faccioli:2010kd}.

%% file: Data.tex
\section[Data analysis]{Data analysis}\label{sec:dataAnalysis}

\subsection[Data set and electron identification]{Data set and electron identification}

The $p+p$ 200 GeV data used in this analysis were recorded by the STAR experiment in the year 2009. 
The STAR detector \cite{Ackermann2003624} is a multi-purpose detector. It consists of many subsystems and has cylindrical geometry and a large acceptance with a full azimuthal coverage. The most important subsystems for this analysis  are briefly described below. The Time Projection Chamber (TPC) \cite{Anderson:2003ur} is the main tracking detector for charged particles. It is also used to identify particles using the ionization energy loss ($dE/dx$). Outside the TPC is the Time Of Flight (TOF) detector \cite{Llope:2012ti} which extends STAR particle identification capabilities to momentum ranges where TPC $dE/dx$ alone is inadequate. Between the TOF and the STAR magnet there is the STAR Barrel Electromagnetic Calorimeter (BEMC) \cite{Beddo:2002zx}. The BEMC is constructed so that an electron should deposit all its energy in the BEMC towers while hadrons usually deposit only a fraction of their energy. The energy deposited by a particle in the BEMC can thus be used to discriminate between electrons and hadrons, by looking at the $E/p$ ratio. The BEMC is also used to trigger on high-$p_{T}$ electrons.
Together with the TOF, the BEMC is utilized to discriminate against pile-up tracks in the TPC, since both detectors are fast. Most of the STAR detector subsystems are enclosed in a room temperature solenoid magnet with a uniform magnetic field of maximum value of 0.5 T \cite{Bergsma2003633}.

The analyzed data were collected with the High Tower (HT) trigger, which requires transverse energy deposited in at least one single tower of the BEMC to be within 2.6 $< E_{T}  \leq$ 4.3 GeV. The HT trigger also requires a coincidence signal from two Vertex Position Detectors \cite{Llope:2003ti}. We have analyzed $\sim 33$ M events with the HT trigger and with a primary vertex $z$ position $|V_{z}| <$ 65 cm. This corresponds to an integrated luminosity of $\sim 1.6$ pb$^{-1}$. The $J/\psi$ is reconstructed via its di-electron decay channel, $J/\psi \rightarrow e^+ e^-$, with the branching ratio $5.94 \% \pm 0.06 \%$ \cite{PhysRevD.86.010001}.

Charged tracks are reconstructed using the STAR TPC which has 2$\pi$ azimuthal coverage and a pseudorapidity ($\eta$) coverage of $|\eta| <$ 1. Tracks that originate from the primary vertex and have a distance of closest approach (DCA) to the primary vertex of less than 2 cm are used. In 2009 STAR did not have a vertex detector that would help to distinguish between prompt and non-prompt $J/\psi$, and TPC resolution alone is not enough to select non-prompt \Jpsi from B meson decays. In order to ensure a good track quality, tracks are required to have at least 15 points used in the track reconstruction in the TPC, and to have at least 52\% of the maximum number of possible track reconstruction points. Cuts of $|\eta| <$ 1 and $p_{T} > $ 0.4 GeV/$c$ are also applied. The transverse momentum cut is chosen to optimize the acceptance in $\cos\theta$ and the significance of the \Jpsi signal. Applying higher $p_{T}$ cut causes a loss  of statistics at $\vert \cos\theta \vert \sim $ 1 while a lower $p_{T}$ cut reduces the \Jpsi signal significance. Efficient identification of electrons with low \pT was possible using available information from the TOF detector. During the analyzed run in 2009, 72\% of the full TOF detector was installed. The TOF pseudorapidity coverage is $|\eta| <$ 0.9. 

In order to identify electrons and reject hadrons, information from the TPC, TOF and BEMC detectors is used.
The TPC provides information about $dE/dx$ of a particle in the detector. Electron candidates are required to have $n\sigma_{\rm electron}$ within $-1 < n\sigma_{\rm electron} < 2$, where $n\sigma_{\rm electron} = \log[(dE/dx)/(dE/dx\mid_{\rm{Bichsel}})]/\sigma_{dE/dx}$, $dE/dx$ is the measured energy loss in the TPC, $dE/dx\mid_{\rm{Bichsel}}$ is the expected value of $dE/dx$ from the Bichsel function prediction \cite{BichselFun} and $\sigma_{dE/dx}$ is the $dE/dx$ resolution. The Bichsel function is used to calculate the energy dependence of the most probable energy loss of the ionization spectrum from a detector. In a thin material such as the TPC gas, it has been shown that the Bichsel function is a very good approximation for the $dE/dx$ curves \cite{Bichsel:2006cs}. At lower momenta (p $\lesssim$ 1.5 GeV/$c$ ), where electron and hadron $dE/dx$ bands overlap, the TOF detector is used to reject slow hadrons. For $p <$ 1.4 GeV/$c$, a cut on the speed of a particle, $\beta$, of $\vert 1/\beta -1 \vert  < $ 0.03 is applied. At higher momenta, the BEMC rejects hadrons efficiently. For momenta above 1.4 GeV/$c$, a cut on $E/p >$ 0.5 $c$ is used for electron identification, where $E$ is the energy deposited in a single BEMC tower ($\Delta \eta \times \Delta \phi = 0.05 \times 0.05$). For electrons, the ratio of total energy deposited in the BEMC to the particle's momentum is expected to be $\approx$ 1. In the analysis we use energy deposited in a single BEMC tower but an electron can deposit its energy in more towers, therefore the value of the $E/p$ cut is 0.5 $c$.

It is also required that at least one of the electrons from the $J/\psi$ decay satisfies the HT trigger conditions. In order to ensure that a selected electron indeed fired the trigger, an additional cut of $p_{T} >$ 2.5 GeV/$c$ is applied for that electron. The HT trigger requirements reduce significantly the combinatorial background under the \Jpsi signal and lead to a clear \Jpsi signal at 2 $< p_{T} <$ 6 GeV/$c$.

\begin{figure}[tbp]
\centering
\includegraphics[width=0.9\columnwidth]{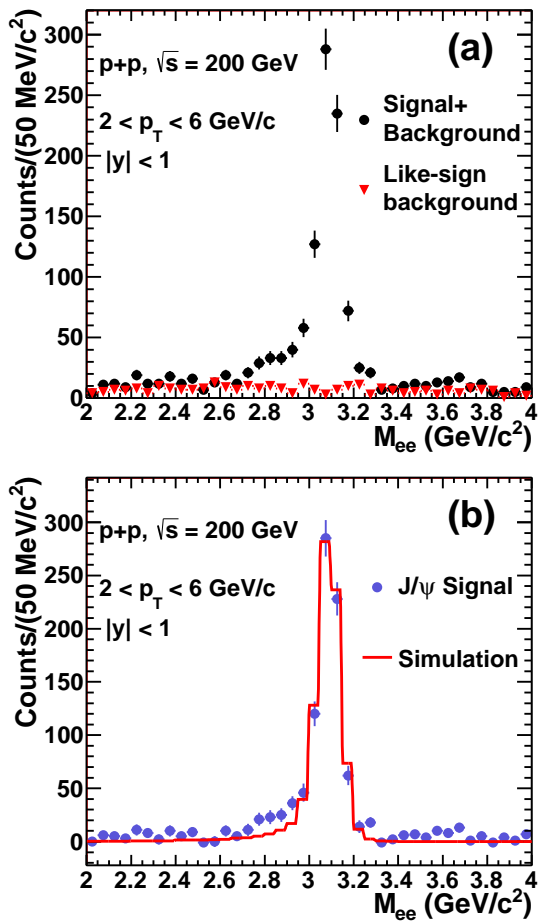}
\vspace{-5pt}
\caption{(Color online.) (a) Invariant mass distributions of unlike-sign (black circles) and like-sign (red triangles) electron/positron pairs, for  2 $<p_{T}<$ 6 GeV/$c$ and $|y| < 1$. (b) \Jpsi signal after the combinatorial background subtraction (closed blue circles) and MC simulation (histogram).}
\label{fig:JpsiMass}
\end{figure}

\subsection[$J/\psi$ signal and $\cos\theta$ distributions]{$J/\psi$ signal and $\cos\theta$ distributions}\label{sec:JpsiSignal}

Electrons and positrons that pass track quality and electron identification (eID) cuts are paired in each event. Figure \ref{fig:JpsiMass}(a) shows the invariant mass distribution for di-electron pairs with $|y| < $ 1 and $p_{T}$ of 2 - 6 GeV/$c$. The unlike-sign pairs are represented by circles. The combinatorial background is estimated using the like-sign technique, and is defined as a sum of all $e^{+}e^{+}$ and $e^{-}e^{-}$ pairs in an event, represented by triangles. The \Jpsi signal is obtained by subtracting the combinatorial background from the unlike-sign pair distribution. Figure \ref{fig:JpsiMass}(b) shows the invariant mass distribution for \Jpsi as circles, and the histogram is the \Jpsi signal obtained from a Monte Carlo (MC) simulation (see Sec.~\ref{sec:corrections}). Momentum resolution of electrons and positrons from the MC simulation is additionally smeared in order that the simulated \Jpsi signal width matches the width of the \Jpsi signal obtained from the data. The simulation does not include the \Jpsi radiative decay channel, $J/\psi \rightarrow e^+ e^- \gamma$ \cite{PhysRevD.86.010001,Adamczyk:2012ey}, leading to the discrepancy between data and simulation for invariant mass $\sim$ 2.7 - 2.9 GeV/$c^{2}$. The tail in the data at low invariant mass is due to electron bremsstrahlung and missing photons in the case of the \Jpsi radiative decay reconstruction. We select \Jpsi candidates in the invariant mass range 2.9 - 3.3 GeV/$c^2$ and so the discrepancy between the data and the simulation for the lower mass range does not influence our result.

In the analyzed ranges of rapidity, $p_{T}$, and invariant mass, the signal to background ratio is 15. A strong \Jpsi signal is seen with a significance of 26 $\sigma$. The number of $J/\psi$, obtained by counting data entries in the \Jpsi mass window, is 791 $\pm$ 30. For the polarization analysis, we split the entire \Jpsi sample into 3 \pT bins with a comparable number of $J/\psi$ in each bin: 2 - 3 GeV/$c$, 3 - 4 GeV/$c$ and 4 - 6 GeV/$c$.

\begin{figure*}[tbp]
\centering
\includegraphics[width=0.9\textwidth]{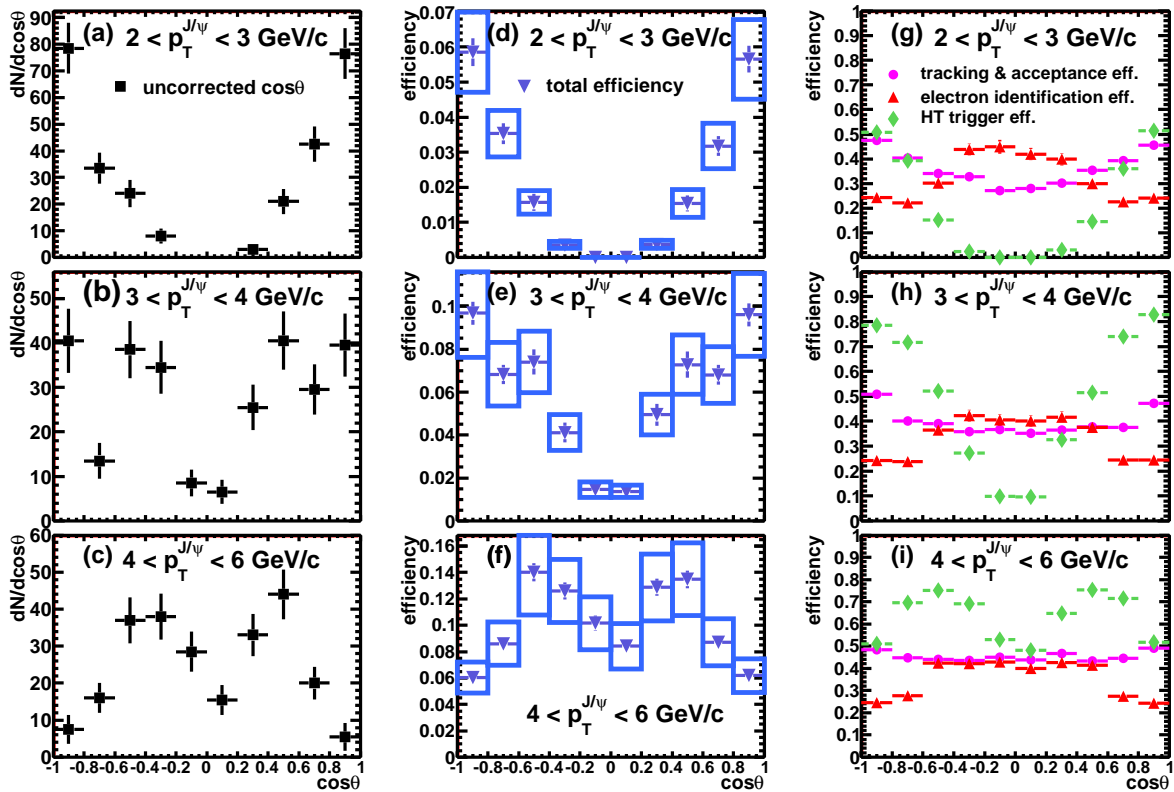}
\caption
{(Color online.) Panels (a)-(c) show uncorrected $\cos\theta$ distributions after the combinatorial background subtraction, for each analyzed \pT bin. Panels (d)-(f) show total efficiencies as a function of $\cos\theta$. Systematic errors are shown as boxes. Panels (g)-(i) show different efficiencies that contribute to the total efficiency.}
\label{fig:Cos_Corr}
\end{figure*}

Raw $\cos\theta$ distributions for $J/\psi$ (after the combinatorial background subtraction) are obtained by bin counting, using distributions from the data. Figures \ref{fig:Cos_Corr}(a)-(c) show uncorrected $\cos\theta$ distributions (full squares).

\subsection[Corrections]{Corrections}\label{sec:corrections}

In order to obtain the $\cos\theta$ corrections, unpolarized Monte Carlo $J/\psi$ particles with uniform $p_T$ and rapidity distributions are embedded into real events, and the STAR detector response is simulated. Since the input $p_T$ and rapidity shapes influence efficiencies, $J/\psi$ distributions are then weighted according to the $J/\psi$ $p_T$ and rapidity shapes observed in the STAR \cite{Adamczyk:2012ey} and PHENIX \cite{Adare:2006kf} experiments. Corrected $\cos\theta$ distributions are obtained by dividing raw $\cos\theta$ distributions by the corrections calculated as a function of $\cos\theta$, in each analyzed $p_{T}$ bin.

Efficiencies as a function of $\cos\theta$ are calculated by applying the same cuts used in the data analysis to the embedding (simulation) sample. Most corrections related to the TPC response, such as the acceptance (with the $p_{T}$ and $\eta$ cuts) and tracking efficiency, and all BEMC efficiencies, are obtained from the simulation. The $n\sigma_{\rm electron}$ and the TOF response are not simulated accurately in embedding. Therefore the $n\sigma_{\rm electron}$ cut and TOF cut efficiencies are calculated using the data. 

For the calculation of the $n\sigma_{\rm electron}$ cut efficiency, the $n\sigma_{\rm electron}$ distribution from the data is approximated with a sum of Gaussian functions (one Gaussian function for electrons and two Gaussian functions for hadrons), in narrow momentum bins. In order to improve the fitting, the TOF and BEMC eID cuts are applied and the position of the Gaussian fit for electrons is constrained using a high-purity (almost 100\%) electron sample obtained by selecting photonic electrons and subtracting a background from like-sign electron pairs. Photonic electrons are produced from photon conversion in the detector material and Dalitz decay of $\pi^{0}$ and $\eta$ mesons. These electrons are isolated using a cut on the invariant mass of a pair of tracks of $m_{e^{-}e^{+}} <$ 100 MeV/$c^{2}$ and additional electron identification cuts: $\vert 1/\beta -1 \vert  < $ 0.03 for $p <$ 1.5 GeV/$c$ and $E/p >$ 0.5 $c$ for momenta above 1.5 GeV/$c$.

TOF matching efficiency is calculated using a low luminosity data sample (with almost no pile-up). Since the TOF detector did not have full coverage in 2009, the TOF matching efficiency is applied in the total efficiency calculation as a function of $\eta$. The efficiency of the $1/\beta$ cut is calculated by using a pure electron sample obtained by selecting photonic electrons with $-0.2 < n\sigma_{e} < 2$ and with the invariant mass of a pair of tracks less than 15 MeV/$c^{2}$. The $1/\beta$ cut efficiency is calculated in narrow momentum bins and then a constant function is fitted to obtain the final $1/\beta$ cut efficiency.

The total \Jpsi efficiency calculations include contributions from the acceptance, the tracking efficiency, the electron identification efficiency, and the HT trigger efficiency, and are shown as a function of $\cos\theta$ in Fig. \ref{fig:Cos_Corr}(d)-(f) (blue triangles). The systematic uncertainties (discussed in subsection \ref{sec:sysErrors}) on the total efficiency are also shown in the figure. The right-hand panels, Fig. \ref{fig:Cos_Corr}(g)-(i), show separately the efficiencies that contribute to the total efficiency.

The most important factor influencing the shape of the total efficiency is the HT trigger efficiency, which is shown as green diamonds in Fig. \ref{fig:Cos_Corr}(g)-(i). At least one of the electrons from the $J/\psi$ decay is required to satisfy the trigger conditions and must have $p_T$ above 2.5 GeV/$c$. Due to the decay kinematics this cut causes significant loss in the number of observed $J/\psi$ at lower $J/\psi$ $p_T$, and the efficiency decreases with decreasing $|\cos\theta|$. 
This pattern is clearly visible in the HT trigger efficiency plot for 2 $< p_{T} <$ 3 GeV/$c$ in Fig. \ref{fig:Cos_Corr}(g), where all entries at $\cos\theta \sim$ 0 are zero. With increasing $J/\psi$ $p_T$, the trigger efficiency increases. Since the trigger has also an upper threshold ($E_{T} \leq$ 4.3 GeV), a decrease of the efficiency at $|\cos\theta| \sim$ 1 at higher \pT is seen, as evident in Fig. \ref{fig:Cos_Corr}(i).

%% file: Result.tex
\section[Results and Discussion]{Results and Discussion}\label{sec:results}

\subsection[Corrected $\cos\theta$ distributions]{Corrected $\cos\theta$ distributions}

\begin{figure}[]
\centering
\includegraphics[width=0.9\columnwidth]{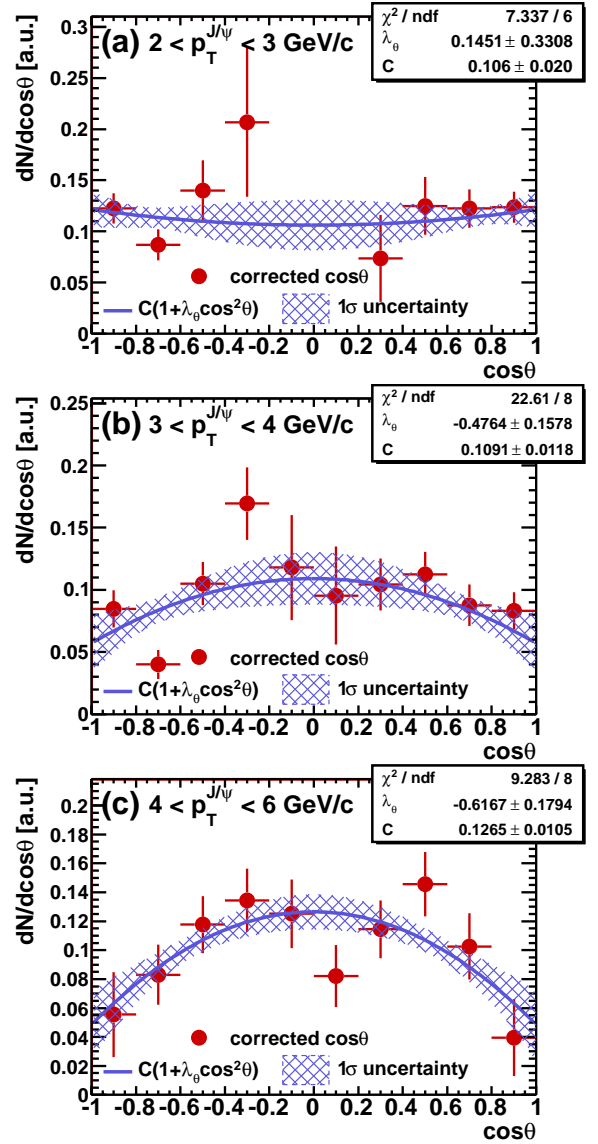}
\centering
\vspace{-5pt}
\caption{(Color online.) Corrected $\cos\theta$ distributions fitted with the function in Eq. \ref{eq:lambdaFit}.  The plotted errors are statistical. The solid blue lines represent the most likely fits, and the hatched blue bands represent the 1$\sigma$ uncertainty on the fits.}
\label{fig:corrCos}
\end{figure}

The corrected $\cos \theta$ distributions are fitted with
\begin{equation}
f(\cos\theta) = {C} (1+\lambda_\theta \cos^2\theta)
\label{eq:lambdaFit}
\end{equation}
where C is a normalization factor and $\lambda_\theta$ is the polarization parameter. The fitting procedure is carried out with no constraints applied to the fit parameters. The corrected $\cos \theta$ distributions with the fits are shown in Fig. \ref{fig:corrCos}. The errors shown are statistical only. The solid line represents the most likely fit. The band around the line is a 1$\sigma$ uncertainty contour on the fit, which takes into account uncertainties on both fit parameters and correlations between them. The measured values of the polarization parameter, in each analyzed \pT bin, are listed in Table \ref{tab:lambda} together with a mean $p_{T}$ ($\langle p_T \rangle$) in each bin and statistical and systematic uncertainties. 

\begin{table}[h]
\centering
\small
\begin{tabular}{c|c|p{4cm}}
$p_{T}$ (GeV/$c$) & $\langle p_T \rangle$ (GeV/$c$) & $\lambda_{\theta}$ \\ \hline \hline
2 $<p_{T}<$ 3 & 2.48 & 0.15 $\pm$ 0.33 (stat.) $\pm$ 0.30 (sys.) \\ \hline
3 $<p_{T}<$ 4 & 3.52 & -0.48  $\pm$ 0.16 (stat.) $\pm$ 0.16 (sys.) \\ \hline
4 $<p_{T}<$ 6 & 4.74 & -0.62 $\pm$ 0.18 (stat.) $\pm$ 0.26 (sys.) \\
\end{tabular}
\caption{The polarization parameter $\lambda_{\theta}$.}
\label{tab:lambda} 
\end{table}

\subsection[Systematic uncertainties]{Systematic uncertainties}\label{sec:sysErrors}

The systematic uncertainties on the polarization parameter $\lambda_{\theta}$ are summarized in Table \ref{tab:sysErrors}. All sources, except the last two, contribute to the error on the total efficiency and are included in the systematic uncertainties shown in Fig. \ref{fig:Cos_Corr}(d)-(f). Each contribution is described below. Each systematic uncertainty is the maximum deviation from the central value of $\lambda_{\theta}$. The systematic uncertainties are combined assuming that they are uncorrelated, and are added in quadrature.

 \vspace{-5pt}
\subsubsection*{Tracking efficiency}
The systematic uncertainty on the tracking efficiency arises from small differences between the simulation of the TPC response in the embedding calculation and the data. Track properties, DCA and the number of points used in the track reconstruction in the TPC (fitPts), are compared between simulation and data. The systematic uncertainty is due to a shift of the fitPts distribution (by 2 points) in the simulation. The uncertainty is considered symmetric.

 \vspace{-5pt}
\subsubsection*{TPC eID efficiency}
The systematic uncertainty from TPC electron identification is estimated by changing constraints on the mean and width of the Gaussian fit for electrons and recalculating the total efficiency. The constraints put on the mean and width are allowed to vary by 3$\sigma$.

 \vspace{-5pt}
\subsubsection*{TOF efficiency}
Since the TOF detector did not have full coverage in 2009, the TOF matching efficiency is applied in the total efficiency calculation as a function of $\eta$. The systematic uncertainty is estimated with the TOF matching efficiency also being a function of azimuthal angle $\phi$. The $1/\beta$ cut efficiency estimated from the data in small $p_{T}$ bins may be sensitive to fluctuations. The $1/\beta$ distribution obtained from the data is well described by the Gaussian function. So the systematic uncertainty on the $1/\beta$ cut efficiency is estimated by applying the efficiency calculated for the whole momentum range of 0.4 $< p < $ 1.4 GeV/$c$ from a Gaussian fit to the $1/\beta$ distribution.

 \vspace{-5pt}
\subsubsection*{BEMC efficiency}
Differences between the simulated BEMC response and the BEMC response in the real data may affect the matching of a TPC track to the BEMC detector and the efficiency of the $E/p$ cut. The matching efficiency of a TPC track to the BEMC and the $E/p$ distribution are compared between data and simulation.
A pure electron sample from the data is obtained by selecting photonic electrons with $-0.2 < n\sigma_{e} < 2$ and with the invariant mass of a pair of tracks less than 15 MeV/$c^{2}$.
The systematic uncertainty of the BEMC efficiency is estimated by applying BEMC matching and $E/p$ cut efficiencies obtained from the data instead of using simulated BEMC response, in the total efficiency calculation. 

 \vspace{-5pt}
\subsubsection*{HT trigger efficiency}
HT trigger response in the simulation, energy in a BEMC tower, is compared with the BEMC response in the data. The systematic uncertainty on the HT trigger efficiency is estimated by varying the trigger turn-on conditions in the simulation by the difference seen between data and simulation, which is 3\%.

 \vspace{-5pt}
\subsubsection*{Input \Jpsi distribution in the simulation}
Since the input \Jpsi transverse momentum and rapidity distributions in the simulation are flat, they need to be weighted with realistic \pT and rapidity spectra. In order to estimate a systematic uncertainty, the \pT and rapidity weights are changed. The \pT weight is varied by changing the ranges in which the Kaplan \cite{Kaplan:1978uq} function is fitted to the \pT spectrum. The weight used for rapidity is obtained by fitting a Gaussian function to the rapidity spectrum, and the systematic uncertainty is estimated by assuming that the rapidity shape is flat at mid-rapidity. 

Also, the \Jpsi particles in the simulation are unpolarized (the input $\cos\theta$ distribution is flat). The acceptance of electron and positron from the \Jpsi decay in the detector depends on the \Jpsi polarization. In order to estimate the effect of the unknown \Jpsi polarization on the acceptance calculation, fully transverse ($\lambda_{\theta} = 1$) and fully longitudinal ($\lambda_{\theta} = -1$) \Jpsi polarization is assumed in the embedding analysis. 
A systematic uncertainty is estimated as a difference between the result obtained with no input \Jpsi polarization and the result when \Jpsi in the simulation is polarized. An average uncertainty from the two input \Jpsi polarizations, longitudinal and transverse, is taken as a systematic uncertainty in this study.

 \vspace{-5pt}
\subsubsection*{Errors from the simulation}
Statistical errors on the total efficiencies, determined using the MC simulation, are included in the systematic uncertainties.

 \vspace{-5pt}
\subsubsection*{Polarization of the continuum background}
In Fig. \ref{fig:JpsiMass}(b), it is seen that there is still some residual continuum background after the combinatorial background subtraction. This background consists of correlated $c\overline{c} \rightarrow e^{+}e^{-}$ and $b\overline{b} \rightarrow e^{+}e^{-}$. The continuum background is about 5 \% of the measured $J/\psi$ in the analyzed invariant mass range. Due to the small statistics of the continuum background, we are not able to estimate a polarization of the correlated background using our data. Instead, we use the value obtained by the PHENIX experiment \cite{PhysRevD.82.012001}. They found that the continuum polarization is between $-0.3$ and $0.3$. We estimate a systematic uncertainty by simulating $\cos\theta$ distributions for the residual background taking two extreme values of $\lambda_{\theta}$: $-0.3$ and $0.3$. Then those $\cos\theta$ distributions are subtracted from the corrected $\cos\theta$ distributions from the data, assuming that the residual background is 5\% of the $J/\psi$ yield, in order to estimate the influence of the continuum background polarization on the measured $\lambda_{\theta}$.

\begin{table}[]
\centering
\small
\begin{tabular}{p{3.cm}|p{1.1cm}|p{1.1cm}|p{1.1cm}}
 & \multicolumn{3}{p{3.3cm}}{Systematic uncertainty on $\lambda_{\theta}$, in $p_{T}$ (GeV/$c$) bins}\\ \hline
Source &2 - 3 &3 - 4 &4 - 6 \\ \hline \hline
Tracking efficiency & 0.024 & 0.009 & 0.008 \\ \hline
TPC eID efficiency & 0.009 & 0.006 & 0.012 \\ \hline
TOF efficiency & 0.057 & 0.018 & 0.014 \\ \hline
BEMC efficiency & 0.035 & 0.024 & 0.068 \\ \hline
HT trigger efficiency& 0.049 & 0.006 & 0.003 \\ \hline 
Input \Jpsi distributions in the simulation & 0.190 & 0.019 & 0.027 \\ \hline
Errors from the simulation & 0.077 & 0.028 & 0.004 \\ \hline
Polarization of the continuum background & 0.025 & 0.034 & 0.034 \\ \hline
$J/\psi$ signal extraction & 0.195 & 0.149 & 0.246 \\ \hline
 \hline
Total & $\pm$0.297 & $\pm$0.160 & $\pm$0.260 \\ 
\end{tabular}
\caption{Systematic uncertainties.}
\label{tab:sysErrors} 
\end{table}

\begin{figure*}[tbp]
\centering
\includegraphics[width=0.65\textwidth]{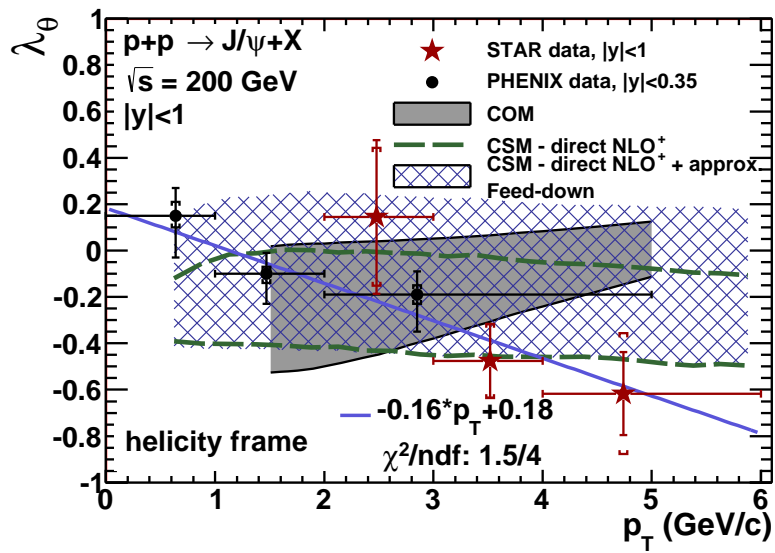}
\caption{(Color online.) Polarization parameter $\lambda_\theta$ as a function of $J/\psi$ $p_T$ (red stars) for $|y|$ $<$ 1.  The data is compared with the PHENIX result (black solid circles) \cite{PhysRevD.82.012001} and two model predictions: NLO$^{+}$ Color Singlet Model (CSM) (green dashed lines represent a range of  $\lambda_{\theta}$ for the direct \Jpsi, and the hatched blue band is an extrapolation of $\lambda_{\theta}$ for the prompt $J/\psi$) \cite{Lansberg:2010vq_paper} and LO NRQCD calculations with color-octet contributions (COM) \cite{PhysRevD.81.014020} (gray shaded area). The $p_{T}$ coverages of the CSM and COM models are $\sim$ 0.6 - 6.0 GeV/$c$ and $\sim$ 1.5 - 5.0 GeV/$c$, respectively. The horizontal error bars represent widths of $p_{T}$ bins. The blue line is a linear fit (${A}x+{B}$) to RHIC points.}
\label{fig:polarization}
\end{figure*}

 \vspace{-5pt}
\subsubsection*{\Jpsi signal extraction}
The systematic uncertainty associated with the \Jpsi signal extraction is estimated by counting the number of \Jpsi particles using the simulated $J/\psi$ signal shape. The \Jpsi signal from the simulation is extracted in each \pT and $\cos\theta$ bin and fitted to the data.

\subsection[Polarization parameter $\lambda_{\theta}$]{Polarization parameter $\lambda_{\theta}$}

Figure \ref{fig:polarization} shows the polarization parameter $\lambda_\theta$ as a function of $J/\psi$ $p_{T}$ for inclusive \Jpsi production. The result includes direct \Jpsi production, as well as \Jpsi from feed-down from heavier charmonium states, $\psi^{'}$ and $\chi_{C}$ (about 40\% of the prompt \Jpsi yield \cite{PhysRevD.81.051502}), as well as from $B$ meson decays (non-prompt $J/\psi$) \cite{Adamczyk:2012ey}. The non-direct \Jpsi production may influence the observed polarization. The STAR result (red stars) is compared with the PHENIX mid-rapidity ($|y| < 0.35$) \Jpsi polarization result for inclusive $J/\psi$ \cite{PhysRevD.82.012001} (black solid circles). The blue line is a linear fit, which takes into account both statistical and systematic uncertainties, to all RHIC points. The fit gives a negative slope parameter $-0.16 \pm 0.07$ with $\chi^{2}$/ndf = 1.5/4. A trend towards longitudinal \Jpsi polarization is seen in the RHIC data. 

STAR observes longitudinal \Jpsi polarization in the helicity frame at \pT $>$ 3 GeV/$c$. The STAR and PHENIX measurements are consistent with each other in the overlapping \pT region. Our result can be compared to the polarization measurements from CDF \cite{PhysRevLett.99.132001} and CMS \cite{Chatrchyan:2013cla} at mid-rapidity for prompt $J/\psi$. At \pT $\sim$ 5 GeV/$c$, CDF observes almost no polarization, $\lambda_{\theta}$ $\sim$ 0 (the polarization becomes slightly longitudinal as \pT increases) while STAR observes a strong longitudinal polarization in that \pT region. At LHC $\sqrt{s}$ = 7 TeV, CMS reports zero polarization at mid-rapidity up to \pT $\sim$ 70 GeV/$c$ \cite{Chatrchyan:2013cla}. In addition, the ALICE experiment also reports very small polarization within 2 $< p_{T} <$ 8 GeV/$c$ at forward rapidity \cite{Abelev:2011md}. However, if the \Jpsi production is $x_{T}$ dependent \cite{Abelev:2009qaa}, the RHIC result at \pT $\sim$ 2 GeV/$c$ is comparable with the CDF result at \pT $\sim$ 20 GeV/$c$ and with the CMS result at \pT $\sim$ 70 GeV/$c$ ($x_{T} \sim$ 0.02, $x_{T} = 2 p_{T}/\sqrt{s}$).

The data are compared with two model predictions for $\lambda_\theta$ at mid-rapidity: NLO$^{+}$ CSM \cite{Lansberg:2010vq_paper} and LO COM \cite{PhysRevD.81.014020}. The prediction of the COM \cite{PhysRevD.81.014020} for direct \Jpsi production, the gray shaded area, moves towards the transverse \Jpsi polarization as \pT increases \cite{PhysRevLett.99.132001}. The trend seen in the STAR and PHENIX results is towards longitudinal \Jpsi polarization with increasing $p_{T}$, and a linear fit to the RHIC data has a negative slope parameter. The difference between the central value of the COM model calculations and the STAR data in terms of $\chi^{2}$/ndf (P value) is 6.7/3 ($8.2 \times 10^{-2}$). The COM failed to describe the polarization measurements by the CDF and CMS experiments at higher energies. 

Green dashed lines represent a range of $\lambda_\theta$ for the direct \Jpsi production from the NLO$^{+}$ CSM prediction and an extrapolation of $\lambda_\theta$ for the prompt $J/\psi$ production is shown as the hatched blue band \cite{Lansberg:2010vq_paper}. It predicts a weak \pt dependence of $\lambda_\theta$, and within the experimental and theoretical uncertainties, the RHIC result is consistent with the NLO$^{+}$ CSM model prediction. Comparison between the central value of the NLO$^{+}$ CSM prediction and the STAR data gives $\chi^{2}$/ndf (P value) of 3.0/3 ($3.9 \times 10^{-1}$) and 5.1/3 ($1.6 \times 10^{-1}$) for the direct and prompt \Jpsi production, respectively.

%% file: Summary.tex
\section[Summary]{Summary and Outlook} 

This paper reports the first STAR measurement of \Jpsi polarization and contributes to the evolving understanding of the \Jpsi production mechanisms. \Jpsi polarization is measured in $p+p$ collisions at $\sqrt{s}$ = 200 GeV in the helicity frame at $|y|<$ 1 and 2 $<p_{T}<$ 6 GeV/$c$. RHIC data indicates a trend towards longitudinal \Jpsi polarization as \pT increases. The result is consistent, within experimental and theoretical uncertainties, with the NLO$^{+}$ CSM model.

Newer data at $\sqrt{s}$ = 500 GeV, taken in 2011 with much higher luminosity, may help to further distinguish between \Jpsi production models, and may permit analysis of the full angular distribution. Furthermore, uncertainties in the models need to be reduced in order to draw more precise conclusions from experimental measurements.

%% file: Acknowledgements.tex
\section*{Acknowledgements}

We thank the RHIC Operations Group and RCF at BNL, the NERSC Center at LBNL, the KISTI Center in Korea and the Open Science Grid consortium for providing resources and support. This work was supported in part by the Offices of NP and HEP within the U.S. DOE Office of Science, the U.S. NSF, CNRS/IN2P3, FAPESP CNPq of Brazil, Ministry of Ed. and Sci. of the Russian Federation, NNSFC, CAS, MoST and MoE of China, the Korean Research Foundation, GA and MSMT of the Czech Republic, FIAS of Germany, DAE, DST, and CSIR of India, National Science Centre of Poland, National Research Foundation (NRF-2012004024), Ministry of Sci., Ed. and Sports of the Rep. of Croatia, and RosAtom of Russia.